\documentclass[3p,twocolumn]{elsarticle}

\usepackage{hyperref}
\usepackage{enumerate} 
\usepackage{amssymb}
\usepackage{algorithmicx,algorithm}
\usepackage{booktabs}
\usepackage{hyperref}
\usepackage{multirow}
\usepackage{stfloats}
\usepackage{graphicx}
\usepackage{color}
\usepackage[justification=centering]{caption}
\usepackage{url}
\usepackage{xcolor}
\usepackage{pdflscape}
\usepackage{amsmath}

\journal{Journal of \LaTeX\ Templates}

\begin{document}

\begin{frontmatter}

\title{AATCT-IDS: A Benchmark Abdominal Adipose Tissue CT Image Dataset for Image Denoising, Semantic Segmentation, and Radiomics Evaluation}

\author[prgA]{Zhiyu Ma}
\author[prgA]{Chen Li\corref{cor1}}
\cortext[cor1]{Corresponding author:} \ead{lichen201096@hotmail.com}
\author[prgA]{Tianming Du}
\author[prgB]{Le Zhang}
\author[prgA]{Dechao Tang}
\author[prgA]{Deguo Ma}
\author[prgA]{Shanchuan Huang}
\author[prgA]{Yan Liu}
\author[prgA]{Yihao Sun}
\author[prgA]{Zhihao Chen}
\author[prgA]{Jin Yuan}
\author[prgA]{Qianqing Nie}
\author[prgC]{Marcin Grzegorzek}
\author[prgB]{Hongzan Sun\corref{cor2}}
\cortext[cor2]{Corresponding author:} \ead{sunhz@sj-hospital.org}

\address[prgA]{Microscopic Image and Medical Image Analysis Group, College of Medicine and Biological Information Engineering,
Northeastern University, Shenyang 110169, China}
\address[prgB]{Shengjing Hospital, China Medical University, Shenyang 110122, China}
\address[prgC]{Institute of Medical Informatics, University of Luebeck, Luebeck, Germany}

\begin{abstract}
  \label{s:abs}

Background and Objective: The metabolic syndrome induced by obesity is closely associated with cardiovascular disease, and the prevalence is increasing globally, year by year. Obesity is a risk marker for detecting this disease. However, current research on computer-aided detection of adipose distribution is hampered by the lack of open-source large abdominal adipose datasets.

Methods: In this study, a benchmark \emph{Abdominal Adipose Tissue CT Image Dataset} (AATTCT-IDS) containing 300 subjects is prepared and published. AATTCT-IDS publics 13,732 raw CT slices, and the researchers individually annotate the subcutaneous and visceral adipose tissue regions of 3,213 of those slices that have the same slice distance to validate denoising methods, train semantic segmentation models, and study radiomics. For different tasks, this paper compares and analyzes the performance of various methods on AATTCT-IDS by combining the visualization results and evaluation data. Thus, verify the research potential of this data set in the above three types of tasks.

Results: In the comparative study of image denoising, algorithms using a smoothing strategy suppress mixed noise at the expense of image details and obtain better evaluation data. Methods such as BM3D preserve the original image structure better, although the evaluation data are slightly lower. The results show significant differences among them. In the comparative study of semantic segmentation of abdominal adipose tissue, the segmentation results of adipose tissue by each model show different structural characteristics. Among them, BiSeNet obtains segmentation results only slightly inferior to U-Net with the shortest training time and effectively separates small and isolated adipose tissue. In addition, the radiomics study based on AATTCT-IDS reveals three adipose distributions in the subject population.

Conclusion: AATTCT-IDS contains the ground truth of adipose tissue regions in abdominal CT slices. This open-source dataset can attract researchers to explore the multi-dimensional characteristics of abdominal adipose tissue and thus help physicians and patients in clinical practice. AATCT-IDS is freely published for non-commercial purpose at: \url{https://figshare.com/articles/dataset/AATTCT-IDS/23807256}.
\end{abstract}


\end{frontmatter}

\section{Introduction}
\label{s:int}

\subsection{Research Background and Motivation}
\label{ss:int-res}

\emph{Metabolic Syndrome} (MetS) is a cluster of metabolic disorders that are becoming increasingly prevalent due to the rising rates of obesity. Each individual component of MetS is a well-established risk factor that is strongly associated with cardiovascular disease (CVD). The obese population tends to exhibit multiple risk factors that increase the incidence and severity of CVD, including heart failure, coronary artery calcification, myocardial infarction, microvascular dysfunction, and cardiac dysfunction~\cite{tune2017cardiovascular}. The global prevalence of MetS is estimated to be around 25\%, and the number of deaths due to obesity has increased by 28.3\% between 1990 and 2015~\cite{saklayen2018global}. Abdominal obesity, which is linked with all components of MetS, has been shown to possess a higher predictive value in determining the susceptibility to CVD than other indicators~{\cite{kumari2019update,wang2007obesity}. The Japanese Metabolic Syndrome Criteria Examination Committee and the International Diabetes Federation have proposed two sets of MetS criteria, both of which consider abdominal obesity as a mandatory risk marker~\cite{yamagishi2017criteria}.

In clinical practice, the assessment of obesity involves various methods, such as waist circumference (WC) and body mass index (BMI), in addition to basic weight measurement~\cite{graffy2016quantification}. BMI is determined by dividing weight (kg) by height (m) squared, with specific thresholds set by the World Health Organization for adults and children~\cite{wang2008will,parikh2007increasing,wang2007obesity}. Conversely, WC thresholds differ based on gender or ethnic group, necessitating the use of a logistic regression model to determine this value. This model connects the WC threshold to cardiovascular disease (CVD) risk, allowing simultaneous acquisition of WC thresholds for MetS risk factors~\cite{zhu2005race}. However, these processes often rely on manual manipulation and are prone to significant errors. Additionally, the application of different thresholds to diverse populations hinders efficient detection. Moreover, current non-imaging biomarkers, including BMI and WC, do not have the ability to characterize or quantify adipose distribution accurately, especially visceral adipose, nor can any of the measured values accurately reflect abdominal obesity alone. Therefore, there is an urgent need to incorporate computerized mechanical aids to achieve a non-invasive diagnosis, optimize the diagnostic process, and improve diagnostic results.

Medical imaging techniques, such as computed tomography (CT), provide a visual representation of the interior of the human body in a non-invasive manner, thus avoiding the controversy caused by metabolic differences across races when WC reflects subcutaneous and visceral fat~\cite{eckel2005metabolic}. Recent studies have shown that combining computer vision techniques and medical imaging can help improve the efficiency and accuracy of clinical diagnosis by radiologists and enable value-added tasks~\cite{pesapane2018Artificial,zhou2021Review}. Segmentation of medical images is a common and vital task in computer-aided diagnostic processes. It involves separating homogeneous regions within the images. One of the most effective approaches for this task is semantic segmentation using deep learning algorithms. These algorithms excel at rapidly and accurately extracting regions of interest (ROI), such as tissues, organs, and lesions, by classifying images at the pixel level~\cite{hesamian2019Deep,shen2017Deep}. Notably, training with two-dimensional slices of abdominal CT images allows the creation of models capable of segmenting three-dimensional body components in abdominal CT scans. These models achieve accuracy levels that either match or exceed those achieved through manual segmentation by experts~\cite{weston2019Automated,litjens2017survey}. Clinically, it is widely accepted that the visceral adipose tissue area in image slices of specific anatomical locations in the abdomen is highly correlated with overall visceral adipose tissue volume and whole-body adipose tissue volume~\cite{shen2004Total,shen2004Visceral}. However, in the study of radiomics, mining higher-order data from digital images is crucial, and the conventional approach is to extract high-level features containing underlying physiological information from ROI~\cite{gillies2016Radiomics}. These high-level features, which are imperceptible to the naked eye, can be employed to develop models for quantifying disease risk and characterizing pathology when integrated with features outside of radiomics~\cite{pesapane2018Artificial,zhou2021Review}. These models can assist radiologists in image analysis and serve as a foundation for supporting diagnostic decisions. The major obstacle for these applications is the lack of adequate large-scale datasets with usable labels~\cite{litjens2017survey}.

This paper presents a benchmark Abdominal Adipose Tissue CT Image Dataset (AATCT-IDS) for image denoising, semantic segmentation, and radiomics evaluation. AATTCT-IDS is a publicly available dataset containing 13,777  images derived from 300 CT scans. According to a certain slice distance, the researchers extract 3,213 slices from the raw CT slices of AATTCT-IDS for manual annotation to identify the subcutaneous and visceral adipose tissue regions and extract a set of multidimensional features. In this study, the images in the dataset were segmented and clustered based on deep learning and machine learning algorithms, and the performance of the different algorithms was evaluated using the corresponding metrics. Furthermore, clinically common mixed noise species are introduced in the images, and the images are restored using various denoising techniques and evaluated. AATCT-IDS is available at the URL: \url{https://figshare.com/articles/dataset/AATTCT-IDS/23807256}.

The main contributions of this paper are as follows:
\begin{itemize}
\item[•] Multiple experts involved in the development of the abdominal adipose tissue CT image dataset.
\item[•] Demonstrate that AATCT-IDS can be used to distinguish the performance of different algorithms in processing CT image denoising, clustering, and semantic segmentation tasks.
\item[•] This abdominal adipose tissue CT image dataset is published as open source for non-commercial purposes.
\end{itemize}

\subsection{Related Work}
\label{ss:int-rel}

In the field of digital image processing for body composition segmentation, a study~\cite{kullberg2007Automated} employs a dataset consisting of abdominal MRI scans of 31 diabetic patients. This dataset contains 16 consecutive axial slices from the 4th to 5th lumbar vertebrae (L4-L5) for each patient. Nonetheless, the study is limited by utilizing annotations generated semi-automatically by a single radiologist as the reference standard. Another study~\cite{makrogiannis2013Computeraided} constructs a novel dataset to evaluate digital image processing and machine learning algorithms. This dataset comprises 168 single-slice abdominal CT scan images obtained from participants in the Baltimore Longitudinal Study of Aging, focusing specifically on the L4-L5 regions of the body. The images have been semi-manually annotated by the investigators. However, due to the limited sample size used for training, the model's ability to generalize in identifying and removing food residues experiences a loss in specificity.

Some studies combining classical machine learning with deep learning employ proprietary datasets. For instance, the Second XiangYa Hospital of Central South University provides abdominal CT scan images of 20 patients for the segmentation and classification of abdominal fat~\cite{wang2020Effective}. These images are manually annotated by radiologists and uniformly resized to $512 \times 512$ pixels. Another study~\cite{devi2022Development} presented a dataset from abdominal MRI images of 65 patients with diabetes and hypercholesterolemia. The researchers apply Gaussian normalization to the initial data and augment it by performing rotations in four directions. Additionally, a study aiming to quantify adipose tissue employs abdominal MRI images from 75 subjects from the First Hospital of Jilin University~\cite{shen2019Automated}. To mitigate the risk of overfitting, the researchers augmented it with polar and affine transformations.

Adipose tissue segmentation based on deep learning has a broader demand for data. A cross-modal segmentation study~\cite{masoudi2020Adipose} uses abdominal CT and abdominal MRI images of 34 National Institutes of Health patients. The study employs patient-level normalization to preprocess the gray values of CT and MRI images so that the histograms have a specific distribution. In addition, the research team individually rectified the MRI images to remove bias fields. To expand the amount of data, methods such as image translation, reflection, and scaling are applied for data augmentation. In another study~\cite{wang2017twostep}, researchers manually annotate CT images of 40 ovarian cancer patients at the Health Science Center of University of Oklahoma for training a model. Similarly, a different study~\cite{shen2023deep} describes a dataset consisting of enhanced CT images from 43 lung cancer patients at the Chinese Academy of Medical Sciences Cancer Hospital, Shenzhen Hospital. In order to improve the ability of the model to segment soft tissue, the CT value of the images is limited to the range of $[–128,150]$. However, the majority of images in the dataset exhibit small visceral fat regions, significantly hampering the model's ability to accurately segment visceral adipose in tested images. Additionally, the limited number of patients in this dataset poses a challenge in demonstrating the model's generalization ability. In another study~\cite{langner2019Fully}, a dataset comprising abdominal MRI images of 45 diabetic patients is used to compare the performance of U-Net and V-Net in segmenting body composition. Researchers employ a lean tissue filter to assist in marking adipose tissue and remove the top 1\% of the highest gray values in the histogram during data preprocessing. A study~\cite{estrada2020FatSegNet} uses the abdominal MRI images of 38 Rhineland subjects to build a dataset for training models. To ensure an accurate evaluation, two experts manually generate the labels without relying on semi-automatic support. Finally, in a study~\cite{koitka2021Fully} conducted by researchers, they curated a dataset comprising 50 patients, with over half of them undergoing screening for oncological indications. The primary purpose of this dataset was to train a semantic segmentation model. However, it's worth noting that the semantic segmentation in this study does not encompass the identification of visceral adipose tissue.

Table~\ref{tab:rel} presents a comprehensive overview of the pertinent dataset details. However, all the aforementioned datasets suffer from a common limitation - a small number of subjects. This becomes particularly challenging due to the significant interindividual variability and the inherent directional characteristics of medical images. Consequently, conventional data augmentation techniques prove inadequate in effectively tackling this concern. On the other hand, the datasets mentioned are non-public, and some are based on scans of patients with non-obesity-related diseases. Therefore, for the advancement of abdominal adipose segmentation research, there is a pressing need for an open-source dataset comprising an ample number of subjects acquired through routine obesity screenings.

\begin{table*}[htbp!]
\centering
\caption{Data usage in image segmentation of abdominal adipose tissue.}
\label{tab:rel}
\resizebox{\textwidth}{!}{
\begin{tabular}{llllllc}
   \toprule
   Aim & Equipment & Amount & Subjects & Type & Method & Reference \\
   \midrule
   \multirow{2}*{Digital image processing} & MRI & 31 patients & Diabetic & Private & Semiautomatic & \cite{kullberg2007Automated} \\
   ~ & CT & 168 images & Healthy & Private & Semiautomatic & \cite{makrogiannis2013Computeraided} \\
   \multirow{3}*{Machine learning} & CT & 20 patients & Sick & Private & Manual & \cite{wang2020Effective} \\
   ~ & MRI & 65 patients & Hypercholesterolemic & Private & Semiautomatic & \cite{devi2022Development} \\
   ~ & MRI & 75 patients & Sick & Private & Semiautomatic & \cite{shen2019Automated} \\
   \multirow{3}*{Deep learning} & CT \& MRI & 34 patients & Sick & Private & Semiautomatic & \cite{masoudi2020Adipose} \\
   ~ & CT & 40 patients & Ovarian cancer & Private & Manual & \cite{wang2017twostep} \\
   ~ & CT & 43 patients & Lung cancer & Private & Manual & \cite{shen2023deep} \\
   ~ & MRI & 45 patients & Diabetic & Private & Semiautomatic & \cite{langner2019Fully} \\
   ~ & MRI & 38 patients & Healthy & Private & Manual & \cite{estrada2020FatSegNet} \\
   ~ & CT & 50 patients & Cancer (50\%) & Private & Manual & \cite{koitka2021Fully} \\
   \bottomrule
\end{tabular}
}
\end{table*}

\subsection{Structure of Paper}
\label{ss:int-str}

This section expounds on the background and motivation for preparing the dataset and the current datasets involved in related research fields. Section~\ref{s:mat} will introduce the preparation process of AATCT-IDS in detail and evaluate the performance of the dataset from multiple perspectives. The evaluation results will be presented in Section~\ref{s:res}. The content of Section~\ref{s:con} will systematically analyze the previous evaluation results. A summary and outlook will be given at the end of the article.

\section{Material and Method}
\label{s:mat}

\subsection{Dataset Preparation}
\label{ss:mat-pre}

AATCT-IDS contains 13,732 abdominal CT images from 300 subjects. The details of the abdominal adipose tissue CT image dataset are described below:
\begin{itemize}
\item (1) Data Source: \\
\textbf{Stage 1:} Shengjing Hospital of China Medical University provides 330 samples saved in DICOM format. Each sample corresponds to the CT scan sequence of a subject, and consecutive slices of L2-L3 sites are selected from the samples by experienced radiologists with a spacing of 1.25 mm between adjacent slices. \\
\textbf{Stage 2:} Under the guidance of radiologists, biomedical researchers at Northeastern University remove substandard data, such as images with visual defects like artifacts (Shown in Figure~\ref{fig:sam} - (a)), and ultimately retain 300 samples to prepare the dataset.

\item (2) Preparation Process:

\textbf{Step 1:} Only image data from DICOM files are extracted for open source to uphold subject privacy and facilitate data integration into diverse deep learning models for training purposes. Given the substantial variation in CT values across fat, muscle, skin, and bone tissues, and considering the acceptable segmentation accuracy achievable with uint8 data type, CT value within the range of $[-1000,+1000]$ is mapped to grayscale values spanning the range of $[0,255]$. Save resultant images in JPEG format for compatibility and accessibility. \\
\textbf{Step 2:} Given that training with sparse annotations can build a model for segmenting dense volume~\cite{cicek20163D} and adjacent slices of abdominal CT are highly similar, images are extracted with a slice spacing of 5 mm from the complete CT slices, and the extracted 3,213 images are combined into sub-datasets for manual labeling of adipose tissue regions. \\
\textbf{Step 3:} Resize all images to $512 \times 512$ pixels by cropping. \\
\textbf{Step 4:} Seven biomedical researchers from Northeastern University and two Shengjing Hospital of China Medical University radiologists manually label subcutaneous and visceral adipose tissue without using semi-automatic support (See Section~\ref{ss:mat-des} for details). \\
\textbf{Step 5:} Three other senior radiologists from Shengjing Hospital of China Medical University further calibrate the labels after unifying the standard. Figure~\ref{fig:flo} illustrates the preparation process of AATCT-IDS.

\item (3) Scanning device: 256 slice spiral CT (Revolution CT, GE Healthcare).

\item (4) Contrast media: Ioversol (320mgI/ml).

\item (5) Annotation software: Labelme.

\item (6) Data format: Both the complete sequence and extracted sub-dataset are saved in JPEG format. Annotations are stored in two forms, the annotations of subcutaneous adipose tissue and visceral adipose tissue are separated and stored as index images in PNG format, and the manually annotated adipose tissue boundary is recorded in JSON format, which is convenient for subsequent expansion or conversion to other formats (See Table~\ref{tab:sca} for details).

\item (7) The Ethical Committee of Northeastern University approved the preparation and study of AATCT-IDS.
\end{itemize}

\begin{figure*}[htbp!]
\centering
\caption{Data preparation workflow of AATCT-IDS.}
\label{fig:flo}
\includegraphics[width=0.95\textwidth]{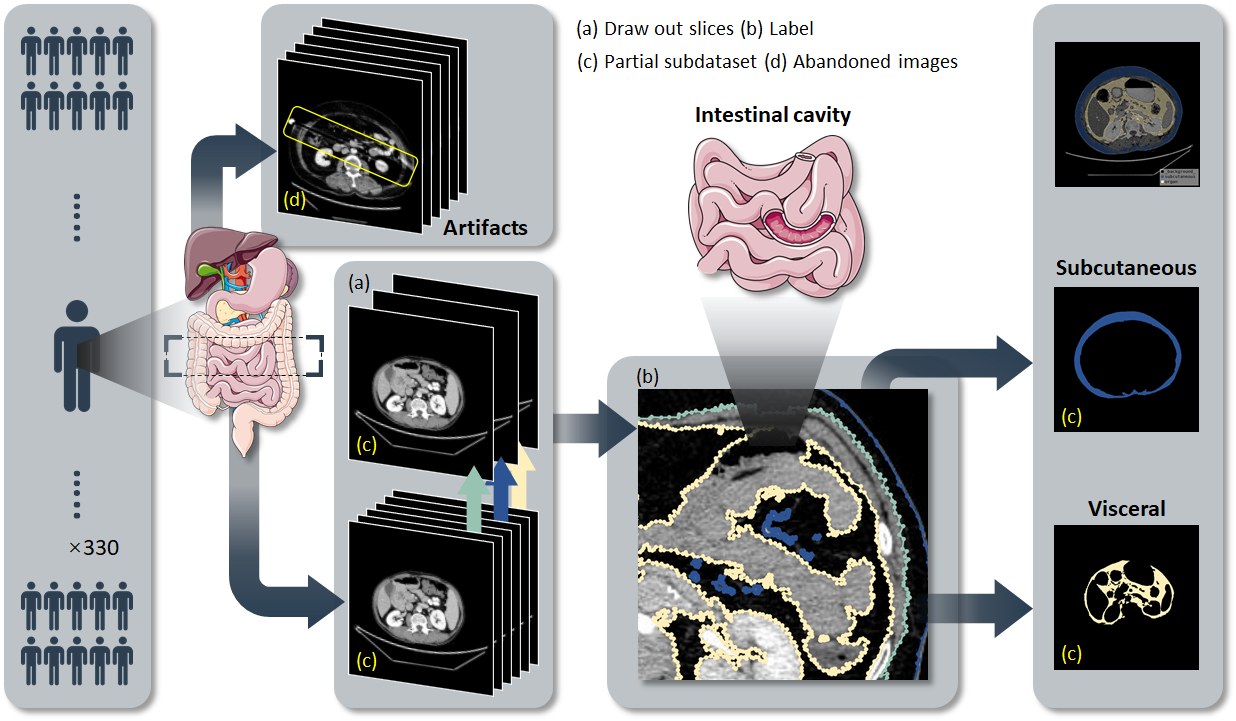}
\end{figure*}

\begin{table*}[htbp!]
\centering
\caption{Dataset scale of AATCT-IDS.}
\label{tab:sca}
\begin{tabular}{llll}
   \toprule
   Aim & Type & Amount & Format \\
   \midrule
   \multirow{4}*{Segmentation} & CT (Partial) & 3,213 & JPEG \\
   ~ & Label (Subcutaneous adipose) & 3,213 & PNG (Index) \\
   ~ & Label (Visceral adipose) & 3,213 & PNG (Index) \\
   ~ & Label (Adipose tissue) & 3,213 & JSON \\
   Denoising \& Radiomics & CT (Total) & 13,732 & JPEG \\
   \bottomrule
\end{tabular}
\end{table*}

\subsection{Dataset Description}
\label{ss:mat-des}

\subsubsection{CT Scan}
\label{sss:mat-des-ct}

Compared to MRI, radiologists can more easily label adipose tissue in CT images~\cite{masoudi2020Adipose}. In addition, CT scans are more readily available and have shorter acquisition times. Thus AATCT-IDS focuses on collecting CT images. This study establishes a set of criteria for volunteer recruitment to ensure a strong association between the data and obesity examination while minimizing confounding variables. The samples in this dataset are derived from a population with randomized BMI, ensuring a balanced representation of both genders. Additionally, the study participants are exclusively individuals who are either healthy or have obesity-related diseases. During the scanning procedure, subjects are instructed to maintain their arms above their heads and follow breath-holding instructions to mitigate the occurrence of artifacts.

\subsubsection{Subcutaneous Adipose Tissue}
\label{sss:mat-des-sub}

CT employs Hounsfield units (HU) to quantify the extent of X-ray beam attenuation within tissues, thereby reflecting tissue density. Adipose tissue has a lower density than muscle, organs, and bone, with CT values typically in the $[-100,-80]$, so adipose appears lower lightness on grayscale images. However, the presence of edema or fatty infiltration can influence radiation attenuation. Edema increases radiation attenuation within subcutaneous adipose tissue, whereas adipose infiltration reduces radiation attenuation in skeletal muscle. As a consequence, the boundaries between subcutaneous adipose and neighboring skeletal muscle become blurred, hindering the image segmentation process. In addition, a small percentage of extremely lean or obese subjects may also show indistinguishable adipose tissue boundaries in CT images. Such images are excluded during dataset preparation (Shown in Figure~\ref{fig:sam} - (b)(c)(d)).

\subsubsection{Visceral Adipose Tissue}
\label{sss:mat-des-vis}

All images in this dataset originate from the L2-L3 segment of the subject, encompassing part of the small intestine's structure (Shown in Figure~\ref{fig:sam} - (e)). Therefore, the presence of the intestinal cavity can often be observed in the images of the dataset. Notably, the intestinal cavity and visceral adipose tissue share close pixel characteristics, and both show low brightness, so the intestinal cavity can easily be misclassified as visceral adipose tissue~\cite{baggerman2021Edema}. In addition, coexistence with intestinal capillaries tissue and various viscera allows for a complex spatial distribution of visceral adipose tissue. As a result, visceral adipose tissue is isolated in numerous narrow spaces with complex and irregular boundaries. The above details are paid attention to during the dataset preparation process.

\begin{figure*}[htbp!]
\centering
\caption{The factors confusing adipose tissue discrimination and the common structures that are easily misclassified.}
\label{fig:sam}
\includegraphics[width=0.95\textwidth]{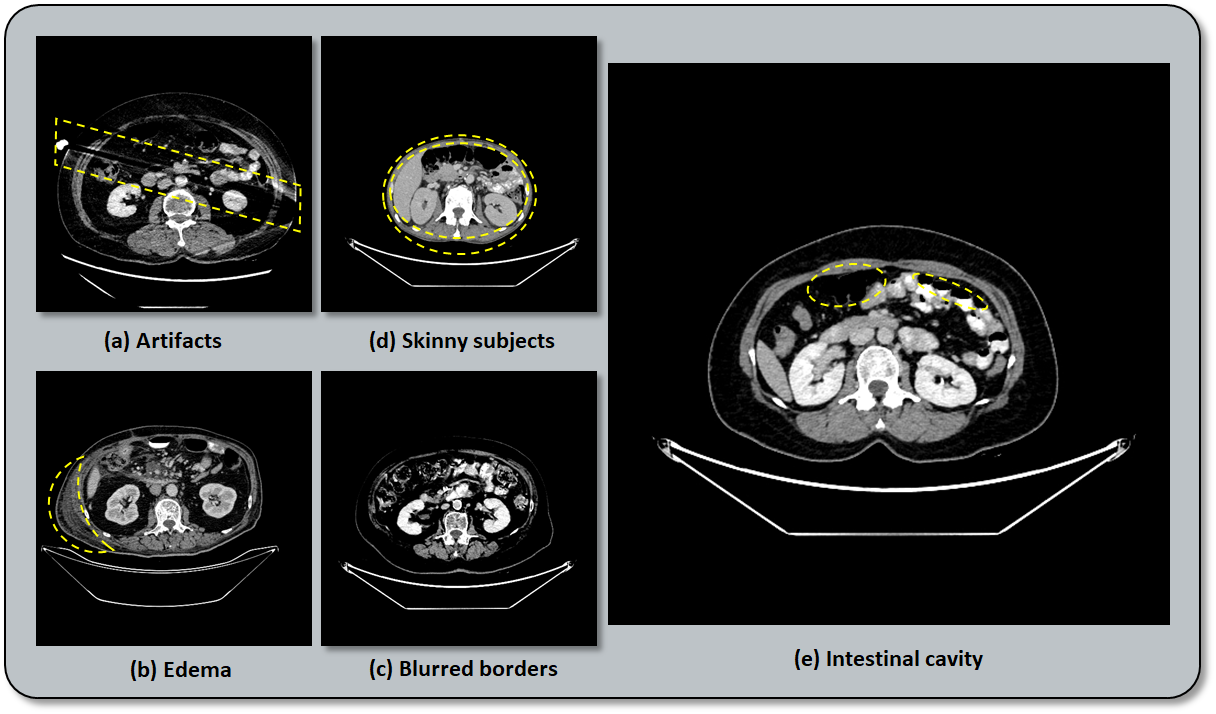}
\end{figure*}

\subsection{Methods of Denoising}
\label{ss:mat-den}

The reconstruction of CT images relies on physical measurements and is sensitive to contrast ratio, so the introduction of noise cannot be avoided~\cite{diwakar2018review}. Denoising while preserving factual medical information is a prerequisite for precision medicine. In this paper, varying degrees of single or mixed types of noise are inserted into all CT images. Subsequently, multiple denoising means are used to restore the image separately, aiming to verify that AATCT-IDS can distinguish the denoising performance of different algorithms.

\subsubsection{Add Noise}
\label{sss:mat-den-add}

Noise in CT images generally follows Poisson and Gaussian distributions, and hardware or software failures can introduce small amounts of noise characterized as Speckle or Pulse types~\cite{ravishankar2017survey}. Poisson noise arises from the inherent uncertainty of photons, while Gaussian noise stems from sensor noise during the acquisition process~\cite{thanh2019review}. In order to simulate the realistic scanning situation, single or mixed types of noise (Gaussian, Poisson, Speckle, Pulse) with a different mean ($\mu=0$, $0.001$) and variance ($\sigma^{2}=0.02$, $0.006$) are introduced into the image.

\subsubsection{Denoising}
\label{sss:mat-den-den}

Image denoising requires restoring the original image as much as possible without destroying the actual image structure, such as boundaries, texture details, and contrast ratio. Denoising algorithms commonly operate in either the spatial or the transform domains, and in some instances, both domains are involved in the denoising process~\cite{kaur2018comprehensive}. In this paper, denoising algorithms based on spatial domain include Median filter, Mean filter, Gaussian filter, and Non-local mean filter (NLM), while denoising algorithms based on transform domain include Wiener filter, Wavelet transform (WT) and Block Matching 3D (BM3D).

The classic smoothing filter is the most basic and popular image denoising method, which can remove some of the low noise with a good time complexity. The Mean filter assigns the arithmetic mean value to the central pixel within a sliding window, effectively reducing Gaussian and Pulse noise. The Median filter is a typical nonlinear filter. During the filtering process, the pixels of the picture will be replaced one by one with the median value of adjacent pixels. In addition, compared with other smoothing filters, it has less damage to edge details. The Gaussian filter is a low-pass filter that blurs the image while reducing noise. The Wiener filter reduces the mean square error of the image by analyzing the statistical characteristics of the image and noise. It exhibits sensitivity to additive noise, making it particularly effective in suppressing Gaussian white noise. The size of the filter window has a direct impact on the denoising effect. Therefore, the smoothing and Wiener filter in this paper applies different filter windows (size$=3$, $5$).

The advantage of the NLM algorithm is that it exploits the redundant information in the image~\cite{buades2005Review}. The replaced gray value in NLM also comes from the weighted average of other pixels. However, unlike the conventional local smoothing filter, the pixels participating in the averaging process are not adjacent but exhibit structural similarity, and the similarity measurement between two pixels is determined based on the Gaussian-weighted Euclidean distance. Furthermore, the filter parameter ($h$) of this algorithm is associated with the attenuation of the exponential function. Excessive values of $h$ can lead to image blurring, while overly small values result in weakened denoising effects. In this paper, the filter parameters of NLM take the same value ($h=\sigma$) as the standard deviation of the noise.

Following a WT, the image is decomposed into components with different frequencies in the wavelet domain. Based on this, the noise can be removed by employing threshold processing at the scale corresponding to each component. The wavelet coefficients associated with noise are eliminated, while those related to structural information are amplified~\cite{mallat1989theory}. In this paper, the decomposition and reconstruction of polluted CT images involve multiple wavelet basis (Daubechies, Dmeyer).

In the denoising process of BM3D~\cite{dabov2007Image}, the algorithm combines the blocks with the minor differences from the reference blocks within a certain distance into a 3D matrix. Subsequently, it sets the smaller coefficients to zero in 3D space by hard thresholding and merges the inverse transformed 3D results to the initial position. It should be noted that a high time complexity accompanies the BM3D algorithm.

\subsection{Methods of Segmentation}
\label{ss:mat-seg}

Training segmentation models is a pre-work of radiomics. Unlike traditional image segmentation algorithms that rely on unsupervised learning, semantic segmentation generates pixel-level predictions on visual inputs and is usually based on deep learning. In this paper, three annotations corresponding to the CT image sub-datasets are used for model training and validation, respectively. All models (FCN, U-Net, PSPNet, DeepLabV3, DenseASPP, BiSeNet, ICNet, DANet) are trained in the same device with the same set of parameters.

By replacing the fully connected layers of traditional CNN, FCN achieves semantic segmentation for the first time in deep learning~\cite{long2015Fully}. The transposed convolution at the end of the FCN uses upsampling to maintain the spatial information of the image, so it is suitable for the input of arbitrary size. In addition, the skip architecture that combines global and local information improves the local prediction based on the global prediction. Therefore, the FCN-8s with relatively accurate segmentation effects are the training object in this paper.

U-Net~\cite{ronneberger2015UNet} and PSPNet~\cite{zhao2017Pyramid} optimize the structure of FCN in different strategies. U-Net pioneered a U-shaped symmetric structure (encoder-decoder), which achieves more thorough feature fusion in a spliced manner. Medical imaging exhibits fixed structure, complex gradients, and small scale. U-Net is widely used in medical image segmentation because it can be trained with smaller datasets while fusing low-resolution information with high-resolution information. PSPNet adds a module called pyramid pooling to the model, which increases the receptive field by paralleling global pooling layers for four scales so that the model can understand the context.

DeepLabV3~\cite{chen2017Rethinking} introduces dilated convolutions in the cascade module, which solves the low-resolution problem caused by consecutive pooling and downsampling. In addition, the improvement of the ASPP module also allows the model to capture sufficient context information. This improvement is achieved by employing stacked or parallel ASPP modules consisting of batch normalization layers and dilated convolutions with varying dilated ratios. Another approach to enhance the ASPP module is demonstrated in DenseASPP~\cite{yang2018DenseASPP}. The non-dense sampling of dilated convolutions may lead to losing much information. Therefore, this model designs a tighter structure splicing dilated convolution for obtaining a larger receptive field.

Both BiSeNet~\cite{yu2018BiSeNet} and ICNet~\cite{zhao2018ICNet} focus on lightweight real-time semantic segmentation tasks. The remarkable feature of the BiSeNet model is the combination of spatial and semantic paths. The shallower networks and more channels of the former are suitable for preserving spatial features, while the deeper networks and fewer channels of the latter are suitable for extracting context. By fusing high-level and low-level features, BiSeNet considers both pixel prediction speed and prediction accuracy. ICNet is a variant of PSPNet. The compressed image is fed into this model from different branches than the original image. The low-resolution image is responsible for providing simple semantic prediction, and the integration of medium-resolution and high-resolution features by the cascade framework can refine the semantic predictions step by step. The performance of ICNet is manifested at less prediction cost.

The novelty of DANet~\cite{fu2019Dual} resides in its integration of attention modules (channel attention module, position attention module) and conventional dilated convolution FCN. This module can obtain the corresponding feature dependencies and update the position feature and channel maps according to the weights. Thus, the model learns better features.

\subsection{Methods of Radiomics}
\label{ss:mat-rad}

In order to verify the potential of AATCT-IDS being used to study radiomics, features in multiple dimensions, including texture, area, and perimeter, are extracted from the image, and unsupervised clustering at the patient level is performed based on the features. In addition, the clustering results are analyzed by evaluation indicators.

$K$-Means is a standard prototype-based clustering algorithm. After initializing the set into $K$ clusters, it will alternately iterate the centroid and clusters until the cluster center changes minimally after iteration. The shape of the clusters can judge the clustering effect. Usually, the spherical cluster symbolizes a better iteration result. DBSCAN is a standard density clustering algorithm that can find the largest set of density-connected samples. By recursively incorporating unvisited samples that meet the specified criteria and are in the neighborhood of visited samples into clusters, DBSCAN enables the expansion of clusters. Therefore, it does not necessitate a preset quantity of clusters and can uncover clusters of diverse geometries. The AGglomerative NESting (AGNES) algorithm employs a bottom-up aggregation strategy. Within the computation procedure of AGNES, each sample is initially treated as an independent cluster, and the algorithm continuously merges the two closest clusters while updating the distance matrix until the preset number of clusters is achieved.

\subsection{Evaluation Metric}
\label{ss:mat-eva}

This study uses visualization and evaluation metrics to examine denoising, segmentation, and clustering methods. The evaluation of denoising methods encompasses indicators such as Structural Similarity (SSIM), Normalized Root Mean Squared Error (NRMSE), and Peak Signal-to-Noise Ratio (PSNR). The evaluation of segmentation methods involves Mean Intersection over Union (MIoU), Mean Pixel Accuracy (MPA), and Dice coefficient. Furthermore, clustering methods are evaluated based on Davies Bouldin Index (DBI) and Dunn Validity Index (DVI).

\subsubsection{Denoising Evaluation Metric}
\label{sss:mat-eva-den}

SSIM measures the similarity between images involving three factors (brightness, contrast, structural information), in which brightness and contrast do not affect structural information. SSIM is defined as:

\begin{equation}
\label{eq:ssim}
\operatorname{SSIM}=\frac{\left(2\mu_{m}\mu_{n}+\left(k_{1}l\right)^{2}\right)\left(2\sigma_{mn}+\left(k_{2}l\right)^{2}\right)}{\left(\mu_{m}^{2}+\mu_{n}^{2}+\left(k_{1}l\right)^{2}\right)\left(\sigma_{m}^{2}+\sigma_{n}^{2}+\left(k_{2}l\right)^{2}\right)}
\end{equation}

Given reference image $m$ and denoised image $n$. In Equation~\ref{eq:ssim}, $l$, $\mu$, and $\sigma^{2}$ correspond to the dynamic range of the gray value, mean, and variance, respectively, and $k_{1}$ and $k_{2}$ are constants. According to the definition of the equation, the value of SSIM is in the $[0,1]$ range and positively correlates with the similarity.

The mean square error (MSE) after the squared root can be normalized to NRMSE. MSE can objectively reflect the difference between pixels, and RMSE reduces the impact of outliers on this basis. PSNR also base on MSE, which can reflect whether the image is distorted by calculating the ratio between practical information and noise. MSE, NRMSE, and PSNR can all be defined mathematically:

\begin{equation}
\label{eq:mse}
MSE=\frac{1}{xy}\sum_{i=0}^{x-1}\sum_{j=0}^{y-1}[m(i,j)-n(i,j)]^{2}
\end{equation}

\begin{equation}
\label{eq:nrmse}
NRMSE=\frac{\sqrt{MSE}}{\bar{m}}
\end{equation}

\begin{equation}
\label{eq:psnr}
PSNR=10\cdot\log_{10}\left(\frac{max_{m}^{2}}{MSE}\right)
\end{equation}

The $x$ and $y$ in Equation~\ref{eq:mse} represent the length and width of the image, respectively. $m(i,j)$ and $n(i,j)$ correspond to the gray value at coordinates $(i,j)$ in image $m$ and image $n$, respectively. And the $\bar{m}$ in Equation~\ref{eq:nrmse} represents the mean value of the reference image. The max$_{m}^{2}$ in Equation~\ref{eq:psnr} is the maximum gray value of the image. Moreover, NRMSE negatively correlates with the denoising effect, and PSNR positively correlates with the denoising effect.

\subsubsection{Segmentation Evaluation Metric}
\label{sss:mat-eva-seg}

MIoU computes the average value of the ratio between the intersection and union of labels and predictions for each class present in the given image. On the other hand, MPA calculates the proportion of pixels in the predicted results that match the label to the total number of predictions. In addition, the Dice coefficient is an assessment tool for measuring the similarity between sets of pixels and exhibits a positive correlation with the intersection over union. These above metrics can be precisely defined using mathematical formulas:

\begin{equation}
\label{eq:miou}
\mathrm{MIoU}=\frac{1}{\mathrm{k}+1}\sum_{\mathrm{i}=0}^{\mathrm{k}}\frac{\mathrm{TP}}{\mathrm{FN}+\mathrm{FP}+\mathrm{TP}}
\end{equation}

\begin{equation}
\label{eq:mpa}
\mathrm{MPA}=\frac{1}{\mathrm{k}+1}\sum_{\mathrm{i}=0}^{\mathrm{k}}\frac{\mathrm{TP}+\mathrm{TN}}{\mathrm{FN}+\mathrm{FP}+\mathrm{TP}+\mathrm{TN}}
\end{equation}

\begin{equation}
\label{eq:dice}
Dice=\frac{2\times TP}{FN+TP+TP+FP}
\end{equation}

The $k$ in Equation~\ref{eq:miou} and Equation~\ref{eq:mpa} represent the number of pixel types in the image except for the background, and the TP, FP, TN, and FN are defined in Equation~\ref{tab:con}.

\begin{table}[htbp!]
\centering
\caption{Confusion Matrix.}
\label{tab:con}
\begin{tabular}{c|c|c}
   \hline
   \multirow{2}*{Ground truth} & \multicolumn{2}{c}{Predict mask} \\
   \cline{2-3}
   ~ & Positive & Negative \\
   \hline
   Positive & TP & FN \\
   \hline
   Negative & FP & TN \\
   \hline
\end{tabular}
\end{table}

\subsubsection{Clustering Evaluation Metric}
\label{sss:mat-eva-clu}

No clinical information involving the privacy of subjects is publicly available in AATCT-IDS, so this paper uses an internal evaluation metric to evaluate the clustering results. DBI reflects the intra-cluster and inter-cluster similarity by calculating the ratio of intra-cluster distance to inter-cluster distance of any two clusters. Therefore, the smaller the metric, the better the clustering performance is characterized. DVI calculates the ratio of the minimum distance (inter-cluster) to the maximum distance (intra-cluster) among all sample points of any two clusters. Both can be defined mathematically:

\begin{equation}
\label{eq:dbi}
DBI=\frac{1}{k}\sum_{i=1}^{k}\max _{j\neq i}\frac{\overline{s_{i}}+\overline{s_{j}}}{\left\|w_{i}-w_{j}\right\|_{2}}
\end{equation}

\begin{equation}
\label{eq:dvi}
DVI=\frac{\min_{1\leq i<j\leq n}d(i,j)}{\max_{1\leq k\leq n}d^{\prime}(k)}
\end{equation}

The aim of Equation~\ref{eq:dbi} is to cluster the samples into $k$ clusters, where $\overline{s_{i}}$ represents the average Euclidean distance from the sample to the center in the $i$ cluster, and ${\left\|w_{i }-w_{j}\right\|_{2}}$ represents the Euclidean distance between the center of the $i$ cluster and $j$ cluster. Furthermore, in Equation~\ref{eq:dvi}, $d(i,j)$ and $d^{\prime}(k)$ represent the Euclidean distance between two samples belonging to different clusters and belonging to the same cluster, respectively.
\section{Result and Analysis}
\label{s:res}

\subsection{Results and Analysis of Denoising Methods}
\label{ss:res-den}

Various denoising methods participate in the comparative study of this paper, and these methods are used to restore the AATCT-IDS data that are contaminated by noise. Figure~\ref{fig:idr} exhibits the visualization outcomes obtain from different combinations of denoising algorithms and noise, all of which correspond to the same original image. Due to space limitations, this figure presents the denoising results for some parameters. Specifically, identical parameters are applied to all types of noise ($\mu=0$, $\sigma^{2}=0.02$), with a filter window size of 3 for both the smoothing filter and the Wiener filter and Daubechies as the wavelet basis for the WT. Detailed multi-parameter results are presented in Table~\ref{tab:den}. Where Gaussian-1 and Gaussian-2 denote Gaussian noise with parameters $\mu=0$, $\sigma^{2}=0.02$ and $\mu=0.001$, $\sigma^{2}=0.06$, respectively. Wavelet-1 and Wavelet-2 indicate that the WT with Daubechies and Dmeyer as wavelet functions, respectively. Also, the sizes in this table mean the filter window size.

This study uses images contaminated with low-intensity Poisson noise, so the relevant results all present nice visual effects. The data in Table~\ref{tab:den} further indicate that the Wiener filter and the NLM algorithm are more specific to this noise. However, the initially complex visceral adipose boundary becomes smooth after NLM processing, which confirms that the noise removal by NLM sacrifices image details. In addition, the median filter with a larger filter window obtains better data results in removing Gaussian noise, but its corresponding image is too blurred. In particular, for some images of thin subjects, the partial outer boundaries of the subcutaneous adipose area are destroyed, which is also reflected in other methods using filter windows.

Mixed types of noise exhibit a more complex distribution, posing a greater challenge for algorithms to suppress the noise effectively. Although the smoothing strategy in this study yields a higher evaluation index for this kind of noise, the resulting images generally fail to maintain adequate sharpness. Conversely, denoising methods based on WT show advantages in restoring AATCT-IDS data contaminated by multiple noises. This method balances the Signal-to-Noise Ratio and image details, achieving comparable objective data to the smoothing strategy while effectively preserving the original image grayscale. In addition, BM3D also preserves most of the image structure, especially when dealing with contamination caused by pulse noise. The excellent performance of BM3D comes at the cost of increased time complexity, and its restoration of AATCT-IDS data consumes much computing time.

\begin{figure*}[htbp!]
\centering
\caption{The result of a specific structure in the image being polluted by different noises and then restored by various denoising algorithms.}
\label{fig:idr}
\includegraphics[width=0.95\textwidth]{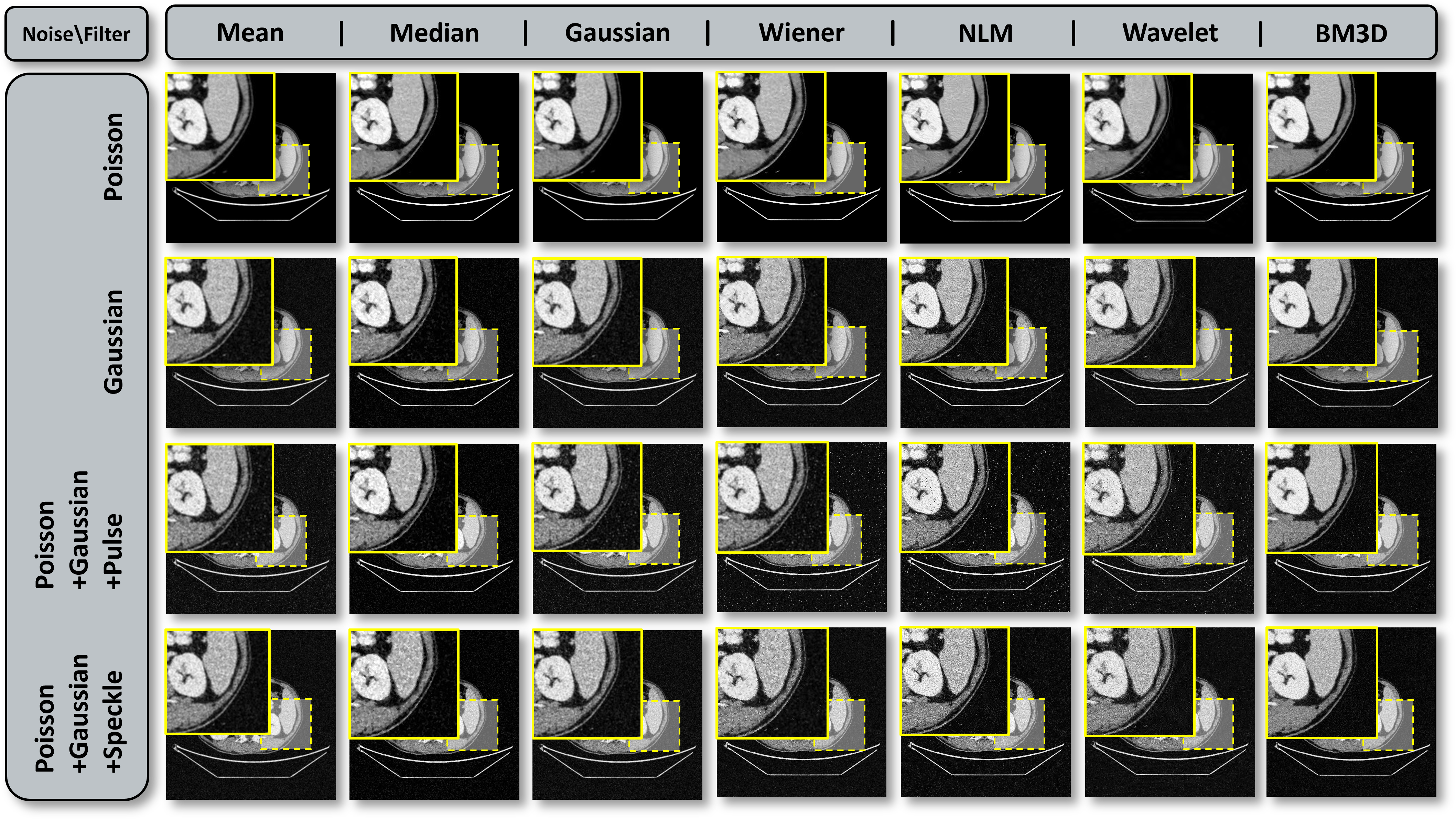}
\end{figure*}

\begin{table*}[htbp!]
\centering
\caption{Results of evaluating the ability of eight denoising algorithms to suppress six types of noise using three metrics.}
\label{tab:den}
\resizebox{\textwidth}{!}{
\begin{tabular}{llllllllllll}
   \hline
   \multirow{2}*{Noises} & \multirow{2}*{Methods} & \multicolumn{3}{l}{Size$=3$} & \multicolumn{3}{l}{Size$=5$} & \multirow{2}*{Methods} & ~ & ~ & ~ \\
   \cmidrule(r){3-5}\cmidrule(r){6-8}
   ~ & ~ & SSIM & NRMSE & PSNR & SSIM & NRMSE & PSNR & ~ & SSIM & NRMSE & PSNR \\
   \hline
   Poisson & Mean & 0.936 & 0.164 & 27.7 & 0.869 & 0.253 & 23.9 & NLM & 0.964 & 0.074 & 34.6 \\
   ~ & Gaussian & 0.954 & 0.133 & 29.5 & 0.927 & 0.176 & 27.1 & BM3D & 0.887 & 0.107 & 31.4 \\
   ~ & Wiener & 0.965 & 0.074 & 34.6 & 0.932 & 0.093 & 32.6 & Wavelet-1 & 0.893 & 0.109 & 31.3 \\
   ~ & Median & 0.951 & 0.100 & 32.0 & 0.898 & 0.186 & 26.7 & Wavelet-2 & 0.872 & 0.105 & 31.6 \\
   Gaussian-1 & Mean & 0.229 & 0.300 & 22.6 & 0.200 & 0.340 & 21.4 & NLM & 0.210 & 0.308 & 22.2 \\
   ~ & Gaussian & 0.237 & 0.290 & 22.9 & 0.232 & 0.295 & 22.6 & BM3D & 0.206 & 0.311 & 22.0 \\
   ~ & Wiener & 0.230 & 0.278 & 23.1 & 0.233 & 0.259 & 23.7 & Wavelet-1 & 0.210 & 0.305 & 22.3 \\
   ~ & Median & 0.298 & 0.215 & 25.4 & 0.384 & 0.227 & 24.9 & Wavelet-2 & 0.211 & 0.303 & 22.4 \\
   Gaussian-2 & Mean & 0.179 & 0.453 & 18.9 & 0.159 & 0.464 & 18.7 & NLM & 0.135 & 0.603 & 16.4 \\
   ~ & Gaussian & 0.183 & 0.455 & 18.9 & 0.185 & 0.443 & 19.1 & BM3D & 0.138 & 0.569 & 17.1 \\
   ~ & Wiener & 0.164 & 0.478 & 18.4 & 0.167 & 0.441 & 19.1 & Wavelet-1 & 0.137 & 0.571 & 16.9 \\
   ~ & Median & 0.204 & 0.337 & 21.4 & 0.270 & 0.291 & 22.7 & Wavelet-2 & 0.138 & 0.571 & 16.9 \\
   Poisson & Mean & 0.227 & 0.297 & 22.6 & 0.194 & 0.341 & 21.4 & NLM & 0.207 & 0.314 & 22.1 \\
   Gaussian-1 & Gaussian & 0.234 & 0.288 & 22.8 & 0.230 & 0.297 & 22.6 & BM3D & 0.206 & 0.317 & 22.0 \\
   ~ & Wiener & 0.226 & 0.284 & 23.0 & 0.230 & 0.266 & 23.5 & Wavelet-1 & 0.207 & 0.310 & 22.2 \\
   ~ & Median & 0.296 & 0.219 & 25.2 & 0.382 & 0.231 & 24.7 & Wavelet-2 & 0.207 & 0.308 & 22.2 \\
   Poisson & Mean & 0.222 & 0.305 & 22.3 & 0.192 & 0.346 & 21.2 & NLM & 0.198 & 0.334 & 21.5 \\
   +Gaussian-1 & Gaussian & 0.229 & 0.296 & 22.6 & 0.226 & 0.304 & 22.4 & BM3D & 0.198 & 0.330 & 21.6 \\
   +Speckle & Wiener & 0.216 & 0.302 & 22.4 & 0.220 & 0.283 & 23.0 & Wavelet-1 & 0.197 & 0.329 & 21.7 \\
   ~ & Median & 0.293 & 0.230 & 24.8 & 0.383 & 0.242 & 24.3 & Wavelet-2 & 0.198 & 0.327 & 21.7 \\
   Poisson & Mean & 0.212 & 0.343 & 21.3 & 0.183 & 0.370 & 20.7 & NLM & 0.185 & 0.456 & 18.8 \\
   +Gaussian-1 & Gaussian & 0.218 & 0.341 & 21.4 & 0.216 & 0.336 & 21.5 & BM3D & 0.187 & 0.437 & 19.8 \\
   +Pulse & Wiener & 0.202 & 0.426 & 19.4 & 0.201 & 0.363 & 20.8 & Wavelet-1 & 0.182 & 0.435 & 19.2 \\
   ~ & Median & 0.291 & 0.224 & 25.0 & 0.378 & 0.235 & 24.6 & Wavelet-2 & 0.183 & 0.435 & 19.3 \\
   \hline
\end{tabular}
}
\end{table*}

\subsection{Results and Analysis of Segmentation Methods}
\label{ss:res-seg}

In this study, the training parameters of all models are unified to obtain data with comparative value. As the segmentation outcomes are solely employed for comparative analyses, these parameters are not tuned to be optimal states. Also, each training consisted of 150 epochs. Figure~\ref{fig:ids} shows the segmentation results of different models for the same abdominal CT slice, where the reference image covered with a mask is marked with the word "label." A more objective evaluation of the outcomes is presented in Table~\ref{tab:sre}. Considering that certain clinical studies only examine specific types of adipose tissue, this study conducted separate training and verification using images of subcutaneous adipose, visceral adipose, and overall abdominal adipose.

As the fundamental model, FCN exhibits limitations in effectively separating the narrow visceral adipose tissue between organs. Nonetheless, the appeal of FCN lies in its lower training cost. Both DenseASPP and DeepLabV3 adopt the strategy of optimizing the ASPP module and show approximate segmentation performance on AATCT-IDS. The difference is that DenseASPP further reduces the training time. The visualization results and evaluation metrics of U-Net surpass those of other models, which confirms its wide application in medical image segmentation. It can separate small and isolated adipose tissues around the kidney while avoiding tiny structures such as small intestinal capillaries. BiSeNet stands out as a lightweight real-time semantic segmentation model, delivering impressive evaluation metrics second only to U-Net for AATCT-IDS segmentation. What sets BiSeNet apart is its ability to achieve these results while consuming the least training time, making it a noteworthy advantage. However, ICNet, another lightweight real-time semantic segmentation model, exhibits a considerable performance gap on this dataset, particularly in coping with the segmentation of subcutaneous adipose tissue. Additionally, given the substantial background area within abdominal CT images, the MPA metrics of each model tend to have approximate values.

\begin{figure*}[htbp!]
\centering
\caption{Segmentation results of adipose tissue in a CT scan slice using eight models trained with the same parameters.}
\label{fig:ids}
\includegraphics[width=0.75\textwidth]{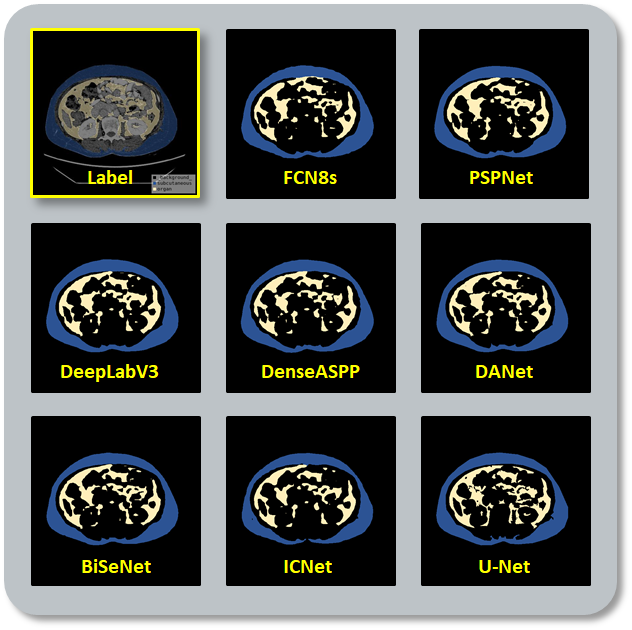}
\end{figure*}

\begin{table*}[htbp!]
\centering
\caption{Results of evaluating the ability of each of the eight models to separate adipose tissue using different metrics.}
\label{tab:sre}
\begin{tabular}{llllll}
   \toprule
   Method & Type & MIoU & MPA & Dice & Training time \\
   \midrule
   FCN8s & Abdominal & 0.828 & 0.960 & 0.888 & 5h5min \\
   ~ & Subcutaneous & 0.889 & 0.984 & 0.908 & 5h8min \\
   ~ & Visceral & 0.831 & 0.962 & 0.884 & 5h6min \\
   PSPNet & Abdominal & 0.855 & 0.968 & 0.903 & 14h6min \\
   ~ & Subcutaneous & 0.910 & 0.987 & 0.924 & 14h19min \\
   ~ & Visceral & 0.847 & 0.969 & 0.897 & 14h22min \\
   DeepLabV3 & Abdominal & 0.849 & 0.971 & 0.900 & 15h29min \\
   ~ & Subcutaneous & 0.900 & 0.990 & 0.918 & 15h43min \\
   ~ & Visceral & 0.853 & 0.972 & 0.901 & 15h42min \\
   DenseASPP & Abdominal & 0.845 & 0.966 & 0.897 & 8h42min \\
   ~ & Subcutaneous & 0.904 & 0.989 & 0.921 & 8h46min \\
   ~ & Visceral & 0.844 & 0.964 & 0.895 & 8h53min \\
   DANet & Abdominal & 0.898 & 0.978 & 0.944 & 14h56min \\
   ~ & Subcutaneous & 0.950 & 0.991 & 0.974 & 15h1min \\
   ~ & Visceral & 0.895 & 0.979 & 0.942 & 15h10min \\
   BiSeNet & Abdominal & 0.901 & 0.978 & 0.947 & 2h37min \\
   ~ & Subcutaneous & 0.957 & 0.991 & 0.978 & 2h36min \\
   ~ & Visceral & 0.900 & 0.976 & 0.945 & 2h43min \\
   ICNet & Abdominal & 0.790 & 0.953 & 0.867 & 4h42min \\
   ~ & Subcutaneous & 0.865 & 0.978 & 0.890 & 4h45min \\
   ~ & Visceral & 0.800 & 0.955 & 0.860 & 4h44min \\
   U-Net & Abdominal & 0.907 & 0.980 & 0.952 & 10h3min \\
   ~ & Subcutaneous & 0.960 & 0.991 & 0.980 & 10h8min \\
   ~ & Visceral & 0.905 & 0.979 & 0.949 & 10h5min \\
   \bottomrule
\end{tabular}
\end{table*}

\subsection{Results and Analysis of Clustering Methods}
\label{ss:res-clu}

Since clustering high-dimensional features is difficult to visualize the results and the evaluation process is complicated, this study uses t-SNE to reduce the multi-dimensional features to two-dimensional space for clustering. In addition, the multi-dimensional features under consideration pertain to the shape or texture features associated with adipose distribution. The outcomes of clustering using $K$-Means on this set of two-dimensional features and the associated evaluation results are presented in Figure~\ref{fig:crk}, where each dot is related to a subject, and dots of the same color are classified as the same cluster. The table on the right side of this figure evaluates the closeness of the different clusters. Moreover, Figure~\ref{fig:crd} and Figure~\ref{fig:cra} demonstrate the clustering and related evaluation results of DBSCAN and AGNES, respectively.

The $K$-Means method in this study performed ten operations, each with a different number of clusters and centroids, and the optimal result is shown in Figure~\ref{fig:crk}. Similarly, the AGNES method undergoes multiple rounds of operations and has the same experimental results as the $K$-Means method, although it defines different orders for the clusters. Both $K$-Means and AGNES clearly classify this set of two-dimensional features into three relatively discrete clusters, which is highly similar to the clustering results of DBSCAN. Therefore, this study divides the subjects into three subsets according to the clustering results of AGNES and conducts a correlation analysis on the shape features and texture features of these three subsets, respectively. The outcomes of this analysis disclosed three distinct patterns of adipose distribution within the dataset population, thereby showcasing the potential of utilizing AATCT-IDS for radiomics research.

\begin{figure*}[htbp!]
\centering
\caption{Results of K-Means clustering subjects based on the features after t-SNE dimensionality reduction.}
\label{fig:crk}
\includegraphics[width=0.95\textwidth]{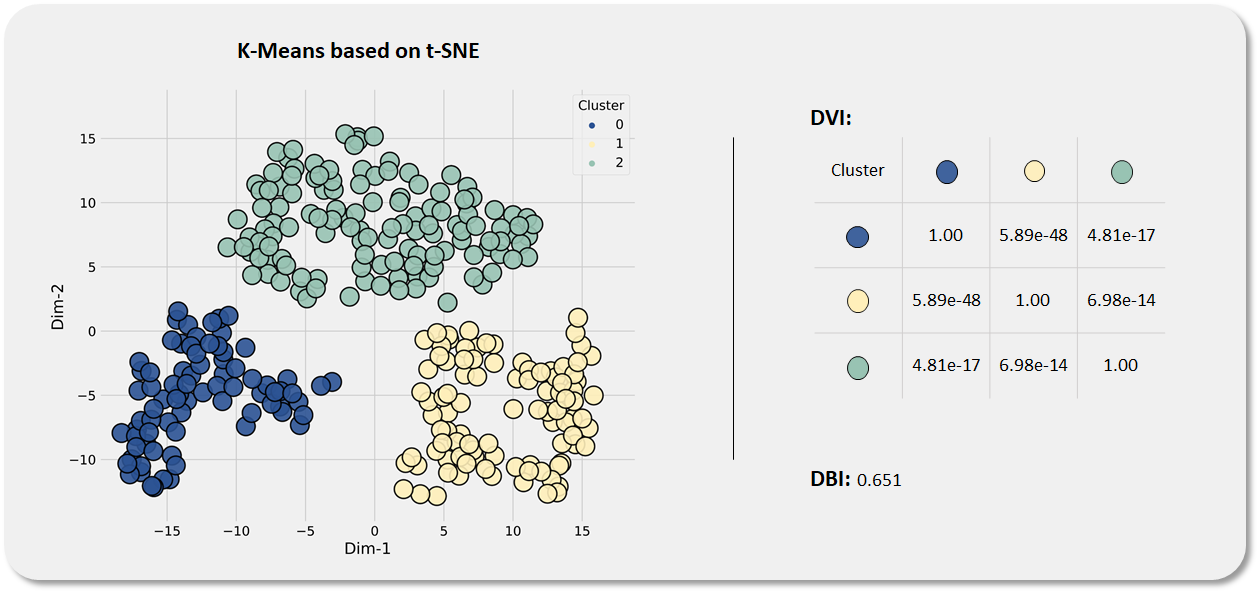}
\end{figure*}

\begin{figure*}[htbp!]
\centering
\caption{Results of DBSCAN clustering subjects based on the features after t-SNE dimensionality reduction.}
\label{fig:crd}
\includegraphics[width=0.95\textwidth]{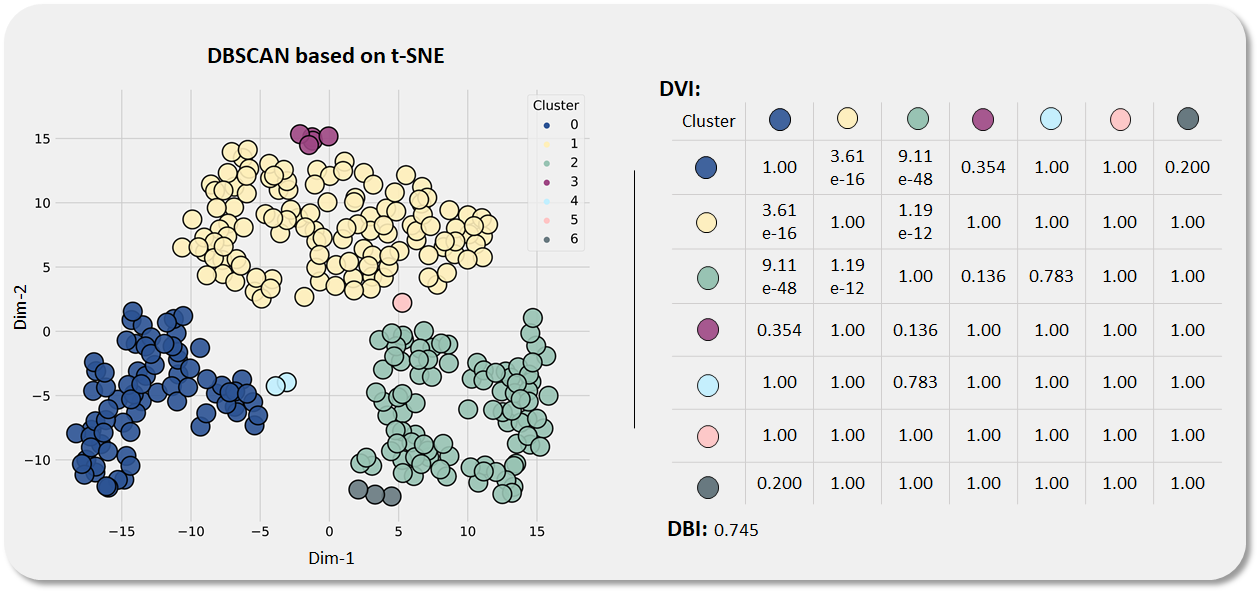}
\end{figure*}

\begin{figure*}[htbp!]
\centering
\caption{Results of AGNES clustering subjects based on the features after t-SNE dimensionality reduction.}
\label{fig:cra}
\includegraphics[width=0.95\textwidth]{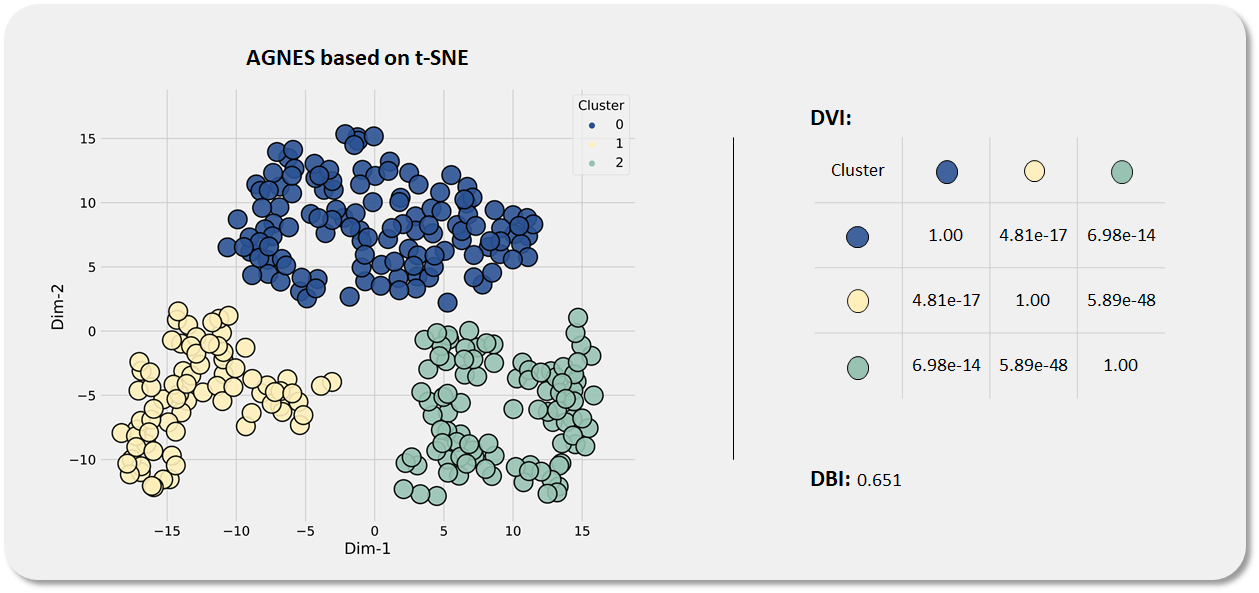}
\end{figure*}

\subsection{Discussion}
\label{ss:res-dis}

Different from conventional natural image segmentation tasks, clinical diagnosis has higher requirements for image prediction accuracy, which is a prerequisite for achieving precision medicine. Some of the segmentation models involved in the comparative study in this paper have obtained good evaluation data. However, many tiny structures in the visualization results of these models still need to make accurate predictions. Therefore, there remains scope for enhancing the accuracy of adipose tissue segmentation in CT images. In addition, abdominal CT images can yield various categories of radiomics features, but only some of the shape and texture features are covered in this paper. Multi-dimensional features containing multiple categories may have a closer relationship with clinical diagnosis.
\section{Conclusion and Future Works}
\label{s:con}

This paper details an abdominal adipose tissue CT image dataset from 300 subjects, AATTCT-IDS. The dataset publics 13,732 raw CT slices, and the researchers individually annotate the subcutaneous and visceral adipose tissue regions of 3,213 of those slices that have the same slice distance. Additionally, all files in the dataset are stored on a subject-by-subject basis.

AATTCT-IDS has the function of examining denoising, segmentation, and radiomics methods. This paper conducts a comparative study on these three types of tasks. In the image denoising study, the CT images in AATTCT-IDS are contaminated with noises of different intensities and distributions. The seven denoising methods in the image restoration step demonstrate different suppression characteristics to these noises. The research delves into image segmentation, employing deep learning to assess the efficacy of eight semantic segmentation models across three clinical medical segmentation tasks. This paper focuses on analyzing the structural characteristics of the adipose tissue segmented by each model while discussing the ability of each model to segment abdominal adipose in combination with the actual training time. In addition, in the radiomics study based on AATTCT-IDS, this paper divides subjects into three subsets by referring to the clustering results of shape features and texture features after dimensionality reduction by various unsupervised methods. Correlation analysis of features performed on these subsets further validates the three fat distributions in AATTCT-IDS.

The comparative analysis of the datasets indicates potential areas for enhancement in the segmentation model. We are committed to exploring a diverse range of approaches for AATTCT-IDS, aiming to unlock a broader scope of possibilities in precision medicine. Furthermore, we intend to integrate clinical diagnostic results with multi-dimensional features spanning various categories, in order to enhance and deepen the study of radiomics.

\section*{Acknowledgments} 

This work is supported by the ``National Natural Science Foundation of China'' 
(No. 82220108007). 
We thank B.A. Qi Qiu, from Foreign Studies College in Northeastern University, 
China, for her professional English proofreading in this paper. 
We also thank Miss. Zixian Li and Mr. Guoxian Li for their important discussion in this work.

\section*{Declaration of Competing Interest} 

The authors declare that they have no conflict of interest.

\section*{Data Availability Statement} 

Ethics is proved by China Medical Universiy, China: No. 2022PS433K. The datasets presented in this study can be found in online repositories. The names of the repository and accession number can be found below: \url{https://figshare.com/articles/dataset/AATTCT-IDS/23807256}.

\bibliography{ma}

\end{document}